# A spintronic Huxley-Hodgkin-analogue neuron implemented with a single magnetic tunnel junction

Davi R. Rodrigues[1],[*] Rayan Moukhader[2,3], Yanxiang Luo[4], Bin Fang[4], Adrien Pontlevy[2], Abbas Hamadeh[5], Zhongming Zeng[4], Mario Carpentieri[1],[*] Giovanni Finocchio[2],[*]

[1]Department of Electrical and Information Engineering, Politecnico di Bari, 70125 Bari, Italy

[2]Department of Mathematical and Computer Sciences, Physical Sciences and Earth Sciences, University of Messina, I-98166, Messina, Italy

[3]Multi-Disciplinary Physics Laboratory, Faculty of Sciences, Lebanese University, Beirut 1500, Lebanon

[4]Key Laboratory of Multifunctional Nanomaterials and Smart Systems, Suzhou Institute of Nano-Tech and Nano-Bionics, CAS, Suzhou, Jiangsu 215123, People's Republic of China

[5]Fachbereich Physik and Landesforschungszentrum OPTIMAS, Technische Universität Kaiserslautern, 67663 Kaiserslautern, Germany

[*]corresponding authors: davi.rodrigues@poliba.it, mario.carpentieri@poliba.it, gfinocchio@unime.it

**Abstract**

Spiking neural networks aim to emulate the brain's properties to achieve similar parallelism and high-processing power. A caveat of these neural networks is the high computational cost to emulate, while current proposals for analogue implementations are energy inefficient and not scalable. We propose a device based on a single magnetic tunnel junction to perform neuron firing for spiking neural networks without the need of any resetting procedure. We leverage two physics, magnetism and thermal effects, to obtain a bio-realistic spiking behavior analogous to the Huxley-Hodgkin model of the neuron. The device is also able to emulate the simpler Leaky-Integrate and Fire model. Numerical simulations using experimental-based parameters demonstrate firing frequency in the MHz to GHz range under constant input at room temperature. The compactness, scalability, low cost, CMOS-compatibility, and power efficiency of magnetic tunnel junctions advocate for their broad use in hardware implementations of spiking neural networks.

I. INTRODUCTION

Spiking Neural Networks (SNNs) are the third generation of artificial neural networks (ANN) and promises significantly reduction of energy consumption, especially for sparse time-dependent data. [1–8] The implementation of biologically-plausible algorithms, which are associated with the low power consumption and the ability to perform non-machine learning computations uphold their applications as robust artificial intelligence accelerators and computing devices while saving battery life. In SNNs, just as in biological brains, information is encoded in the relative time between asynchronous spiking events, which creates collective behavior in a topologically complex network of neurons. The sparse number of events, characterized by high information content, is responsible for the significantly reduced energy consumption and efficient response to event-based excitations. So far, SNN has achieved remarkable success in the domains of visual processing [9,10], speech recognition [11], and medical diagnosis [12].

Scalable SNNs that have some degree of stochasticity competitively emulate brain functions due to their highly parallel operation and in-situ memory and processing. [4–6] Software simulations of bio-realistic neuron synapses and firing are not competitive because they suffer from a high memory and processing cost, even though they are able to precisely emulate the current neuron models. [9,13–18] Hardware implementations of SNNs have been proposed to overcome this obstacle. [19] They significantly increase the efficiency of the network by exploiting optimized task-specific components that mimic the brain functionalities. [4,8,20,21] However, current implementation proposals are unable to optimally fulfill the requirements of SNNs. [4,7,16,22] Current neuromorphic systems, such as IBM's TrueNorth [23] and Intel's Loihi [24] exploit conventional complementary metal-oxide-semiconductor (CMOS) technology and relies mostly on architecture improvements to increase the neural network's efficiency. [25] Implementation of different materials and devices, nonetheless, will allow for a better integration of the necessary neuromorphic properties directly at the material/device level by leveraging their non-linear functional response. [1,26–29] In particular, compact task-oriented devices can substitute complex CMOS circuits to increase scalability and energy-efficiency. Memristors, for example, have gathered great attention due to their potential ability to mimic the spiking behavior as well as synaptic plasticity. [30–34] A promising alternative is the use of spintronic devices which are characterized by low energy dissipation, non-volatility, high speed, reduced sizes, etc. [1,35–40] The demonstrated integration with CMOS-based architectures and the advances in modeling tools of spintronic-CMOS hybrid systems [41] allows for the proposal of optimized spintronic components that can significantly increase area, energy and memory efficiency of current hardware implementations of neural networks. [28,29,42,43] In particular, Magnetic

Tunnel Junctions (MTJs) have already shown to mimic synapse plasticity and firing. [1,37–39,44–49]

Recently, MTJ-based devices have been proposed to emulate the Leaky-Integrate-and-Fire (LIF) model of neurons which has low computational costs. [38,44,45] These MTJ-based proposals rely solely on the magnetization dynamics generated by a series of pulses, which when integrated over time produces a single switch of the magnetization. A common caveat of these proposals is the requirement of a reset current pulse in the opposite direction to return the MTJ to the initial condition. The need of a reset mechanism also implies a clocking system which significantly reduces the computational speed. To overcome the need of a resetting system, it has been proposed the use of MTJs in the stochastic regime, where the magnetization switches between the maximum and minimal resistance configuration due to thermal fluctuations. [50,51] This paradigm suffers from the fact that true spikes are not obtained, since the switch of the magnetization is not immediately followed by the reverse switching. Moreover, the highly stochastic nature of these devices leads to a high error rate which significantly increases the computational time. Other proposals that do not require a resetting mechanism involve the use of antiferromagnetic materials. [35,36] These proposals, however, suffer from the need of a single domain ground state and, in the case of true sharp spikes, the presence of an applied alternate current [35]. A recent spiking MTJ-based proposal has shown that combining the stochastic behavior and ferro-antiferromagnetic coupling allows for the emulation of spiking with single pulses without the need for a resetting pulse. [52] Despite these advances towards the hardware implementation of the computationally cheap LIF model, a scalable room-temperature device that emulates more bio-realistic neuron models is still lacking.

Bio-realistic neuron models, such as the Huxley-Hodgkin (H-H) model [30,33,53–56], exhibit firing at a constant rate for a constant input, called tonic spiking, and allow for spike time-dependent plasticity (STDP) [57–59] and other bio-realistic learning techniques [56,60]. The H-H model embodies the higher efficiency of the brain, and the firing behavior derives from the combination of the different response times of physical processes in the neuron. In this manuscript, we propose a single MTJ device that emulates bio-realistic neurons that fire at frequencies in the MHz to GHz range by leveraging thermal effects and Joule heating. The device produces sharp firing signals followed by a refractory period, which is essential for the implementation of bio-realistic learning processes such as STDP. For a comparison between the H-H model and the behavior of the single MTJ device proposed here, see Table 1. We show that with a single constant input the device at room temperature fires at constant rate and presents a small, but not vanishing, stochasticity due to the thermal field. [61] A single firing response can also be obtained by integrating discrete pulses and thus the device also emulates the LIF model. The proposed device employs a single MTJ, which renders it CMOS-compatible, compact, robust, scalable, and reproducible as required for SNNs. The device design proposed here is based on the idea of hybrid MTJs having the anisotropy easy-axis of the free layer perpendicular to the in-plane polarizer. The main advantage of this solution is the ultralow current working regime and self-oscillation at zero external field as already demonstrated experimentally. [62]

|  | **H-H model** | **Single MTJ device** |
|---|---|---|
| **Output** | In terms of membrane potential | In terms of electrical resistance of device |
| **Physical mechanisms** | Interplay between the concentrations of $Na^+$, $K^+$, and leakage currents | Interplay between magnetization dynamics and thermal effects |
| **Neuron behavior** | Tonic spiking of the voltage at constant input current | Tonic spiking of the voltage at constant input current |
| **Firing rates** | 1-10 KHz [55,63–66] | 100 MHz - 5 GHz |
| **Applied voltage** | 1-200 mV [55,66,67] | 100-300 mV |
| **Refractory period** | Present | Present |

**Table 1.** A comparison between H-H model and the single MTJ device. The performance of the single MTJ device is based on experimental results. [62]

The proposed device concept is presented in detail as follows: In Sec. II, we describe the device properties and model. In Sec. III, we show results obtained from micromagnetic simulations. In Sec. IV, we compare the behavior of the device and the H-H model. In Sec. V, we demonstrate the efficiency of the device in an SNNs considering rate-based information encoding. And finally, Sec. VI provides a summary and outlook.

II. DEVICE AND MODEL

The overall design of the proposed device is shown in Fig. 1(a). Like a biological neuron, the device generates voltage spikes due to an applied current corresponding to the weighted sum of all signals received from other neurons. However, while the variation of the membrane potential in the biological neuron is due the Sodium ($Na^+$), Potassium ($K^+$), and leakage currents, the variation of the potential through the MTJ corresponds to a change in the resistance and is induced by the input current and thermal effects. The active part of the MTJ proposed here has a hybrid configuration where the free layer equilibrium configuration of the magnetization is out-of-plane, and the polarizer is in-plane. As discussed extensively in the experimental work [62], this MTJ configuration has a distinct behavior compared to MTJs where the easy-axis of the free-layer is along the direction of the magnetization in the fixed layer. In this case, the auto-oscillation generated by the STTs induces an oscillation of the resistance since the free-layer magnetization rotates between a state of maximum and minimum resistance, i.e., antiparallel and parallel to the polarizer respectively. The resistance variation in an MTJ generates the spiking behaviour described in Fig. 1(b), which is characterized by a sharp change in resistance followed by a refractory period where the resistance reaches the highest values. Experiments have demonstrated that this MTJ concept exhibits ultralow current density threshold ($<10^6$ A/cm$^2$) for the excitation of the auto-oscillation. [62]

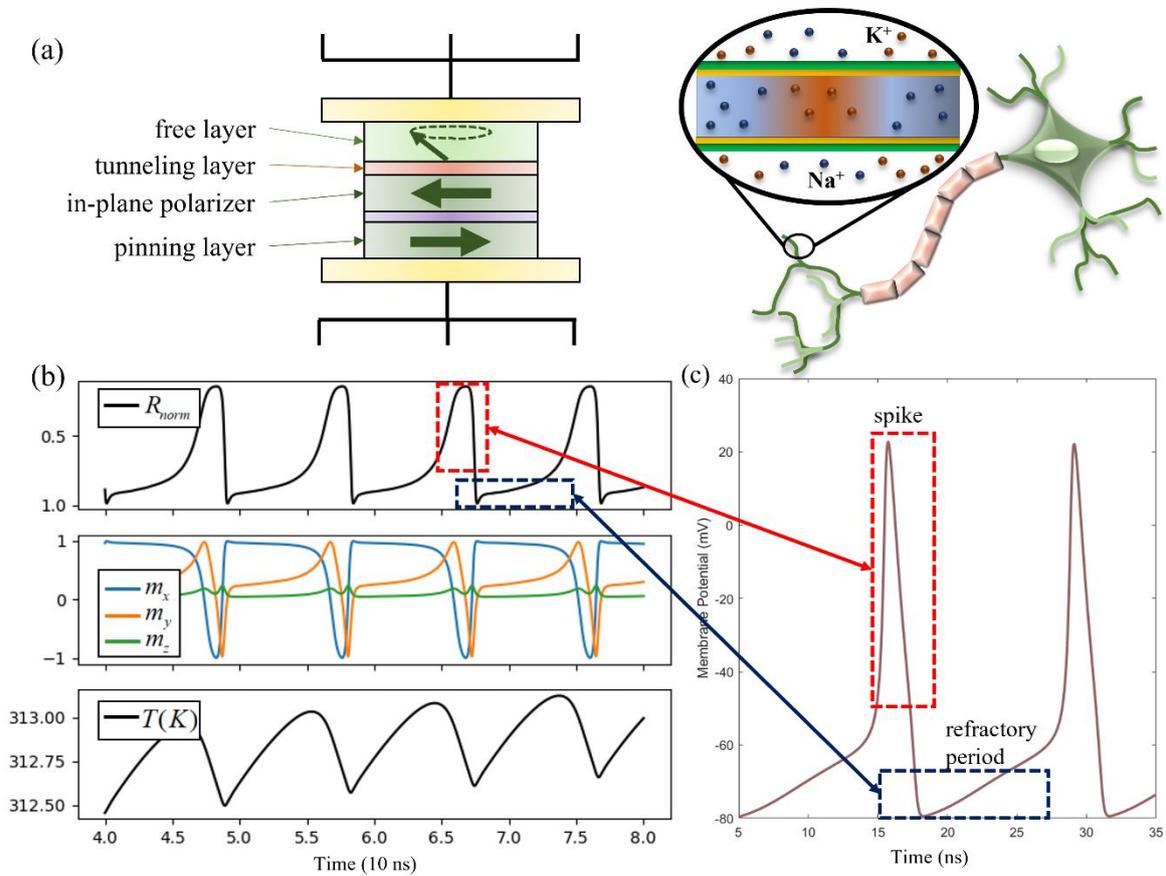

**Figure 1.** An MTJ implementation of bio-realistic firing behavior. (a) On the left we show a sketch of the device concept of the proposed MTJ synaptic device and on the right a sketch of a biological neuron. Both receive the input in terms of currents and produce an output in terms of a voltage variation. In the spintronic device, the potential variation is due to changes in resistance produced by the input current and thermal effects. In the neuron, the potential variation is due to the Sodium (Na$^+$), Potassium (K$^+$), and leakage currents. (b) and (c) compares the resistance variation of the proposed MTJ device with the potential spiking in a biological neuron according to the Huxley-Hodgkin model. In (b) from top to bottom we show: the spikes in the resistance through the device; the dynamics of the magnetization components; the temperature variation of the device.

(c) An example of a numerical simulation of the dynamics of a H-H neuron. It can be observed that the behavior of a sharp firing signal followed by a refractory period is a common feature in both models. For a comparison between the two neurons, see Table 1.

The ultralow threshold is achieved by considering a polarizer composed of a synthetic antiferromagnet that allows for proper control of dipolar fields which drive a small tilting of the equilibrium magnetization of the free layer at zero external field. The relative resistance of the MTJ is given by

$$R_{norm} = \frac{1+P^2}{1-P^2(\mathbf{m}\cdot\mathbf{p})}, \tag{1}$$

where $\mathbf{m} = \mathbf{M}/M_s$ is unitary magnetization vector of the free layer, $M_s$ is the magnetization saturation, $P<1$ is the polarization and $\mathbf{p}$ is the direction of the spin-polarized current. Hence in the absence of spin-transfer torque (STT) [68–71], the resistance of the MTJ is not at a minimal ($\mathbf{m}$ and $\mathbf{p}$ in the same direction) or maximum ($\mathbf{m}$ and $\mathbf{p}$ in opposite directions) state, since the magnetization of the free layer is perpendicular to the magnetization of the polarizer ($\mathbf{m}\cdot\mathbf{p}\approx 0$). When a polarized spin-current along the maximum resistance direction is applied, the STT drives a large amplitude magnetization precession.

To couple the magnetization dynamics with thermal effects, we consider three phenomena:

1) *Resistance variation driven gain-loss*. The first is associated to the gain-loss of temperature of the MTJ driven by variations of the resistance,

$$\frac{dT}{dt} = \rho\frac{R_{norm}(JA)^2}{k_B} - \frac{(T-T_{amb})}{\tau_0}, \tag{2}$$

Where $T$ is the temperature of the MTJ, $\rho$ is the heating efficiency, $J$ is the applied current density, $A$ is the area of the MTJ, $k_B$ is the Boltzmann constant, $T_{amb}$ is the room temperature, and $\tau_0$ the natural temperature decay time of the MTJ. The coefficient $\rho = R_0 t_b / 2\lambda$ depends on the maximum resistance of the device $R_0$, the thickness $t_b$ and thermal conductivity $\lambda$ of the free layer.

2) *Parameter scaling*. The second phenomenon is given by the temperature scaling of micromagnetic parameters, such as the magnetization saturation of the free layer [72] and the polarization [72–75],

$$M_s(T) = M_{s,0}\left(1-\left(T/T_c\right)^{1.5}\right), \tag{3a}$$

$$\text{and } P(T) = P_0\left(M_s(T)/M_{s,0}\right)^{\varepsilon_P}, \tag{3b}$$

where $T_c$ is the Curie temperature, $M_{s,0}$ and $P_0$ are the magnetization saturation and polarization, respectively, at $T=0K$, and $\varepsilon_P$ is a scaling coefficient.

3) *Stochasticity.* The third phenomenon is the thermal excitation of the magnetization in the free layer, which produces a stochasticity of the magnetization trajectory. The thermal excitation is incorporated as a temperature induced white noise, with magnitude [61,76]

$$h_{\text{ther}} = \sqrt{\frac{2k_B \alpha T}{M_s \gamma V_{FL} \Delta t}}, \qquad (4)$$

where $\alpha$ the phenomenological Gilbert damping, $\gamma$ is the gyromagnetic ratio, $V_{FL}$ is the volume of the free layer and $\Delta t$ is the simulation time step.

The first phenomena describe the time-evolution of the temperature which, due to the second phenomena, defines the time evolution of the material parameters. This combined physics is essential to ensure the transient stability of the magnetization in the higher resistivity state and is thus crucial for the characteristic sharp spiking behavior. The third phenomenon generates the stochasticity inherent to bio-realistic neural computation and beneficial to neural network algorithms. [1,3,4,77] The role of each behavior is detailed in the next section.

To describe the magnetization dynamics, we consider a macrospin model based on the Landau-Lifshitz-Gilbert (LLG) equation in the presence of spin-orbit torques, [69,78,79]

$$\frac{d\mathbf{m}}{dt} = -\frac{\gamma}{1+\alpha^2}\left(\mathbf{m} \times (\mathbf{h}_{\text{eff}} + \mathbf{h}_{\text{ther}}) + \alpha \mathbf{m} \times (\mathbf{m} \times \mathbf{h}_{\text{eff}})\right) - \tau_{\text{STT}} \mathbf{m} \times (\mathbf{m} \times \mathbf{p}). \qquad (5)$$

Here $\mathbf{h}_{\text{eff}}$ is the effective field, $\mathbf{h}_{\text{ther}}$ is the thermal field with magnitude given by Eq. 4, and $\tau_{\text{STT}}$ is the spin-transfer torque magnitude given by [62,69]

$$\tau_{\text{STT}} = \frac{\gamma \hbar J}{2M_s e d (1+\alpha^2)}\left(\frac{2P}{1-P^2 \cos\varphi}\right), \qquad (6)$$

where $\hbar/2$ is the magnitude of the electron spin, $J$ is the electric current density, $P$ is the polarization, $e$ is the electron charge, and $d$ is the thickness of the free layer. Based on this set of equations, we can map the dynamics of the biological neuron to the dynamics of the magnetization in the free layer of the proposed single MTJ device. The effective field $\mathbf{h}_{\text{eff}}$ depends on the effective anisotropy vector $\mathbf{D}$ [80], and the dipolar field from the fixed layer $\mathbf{H}_F$. We emphasize that in order to obtain the spiking behavior, it is essential to consider an MTJ with an elliptical cross section as in Ref. [62]. The elliptical shape breaks the in-plane rotational symmetry and favors a single in-plane direction and produces an in-plane unidirectional field $\mathbf{H}_F$ from the polarizer. For details on macromagnetic simulations, see Appendix A. In the next section we describe the results of the macrospin simulations.

III. RESULTS

Fig. 2 shows the change in average resistance and average temperature for different applied currents. In Fig. 2(a), we characterize regions with two distinctive behaviors: (i) and (iii) where the magnetization is fixed by a strong bias given by the anisotropy or the STT, and (ii) where large-amplitude auto-oscillations are excited. Notice that the variation of temperature is essential for the firing behavior, characterized by a switch between steady and auto-oscillating states. The alternation is due to the parameter scaling and the stochasticity as seen on the bottom of Fig 2(a). Fig. 2(b) shows the average temperature in terms of applied current which is responsible for the parameter scaling and the amplitude of the thermally induced stochasticity. Moreover, we emphasize that the stochasticity of the device due to the thermal fluctuations also induce a small level of stochasticity of the firing behavior which is relevant for neuromorphic applications.

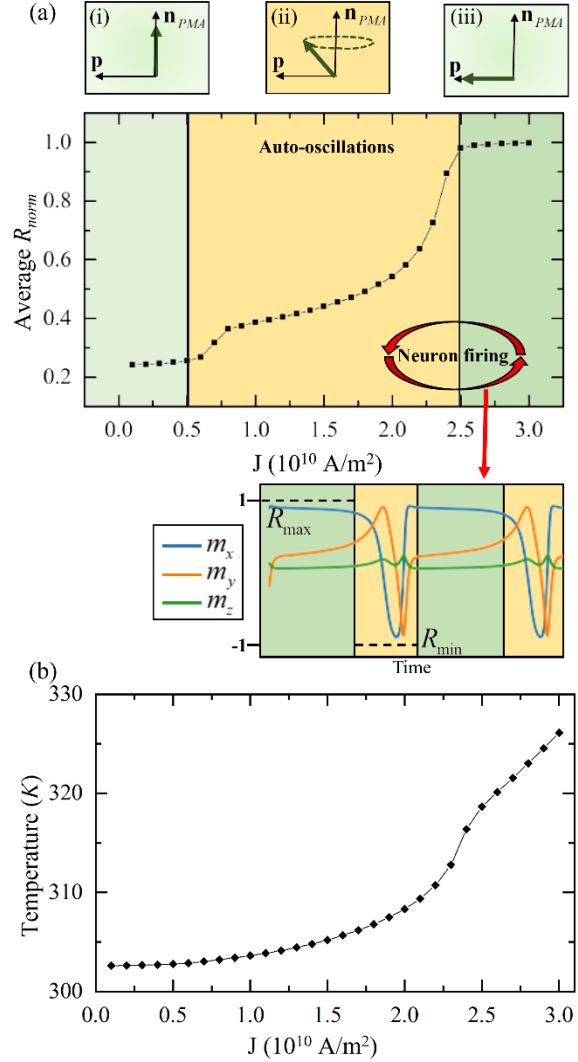

**Figure 2.** (a) and (b) show the dependence of the average normalized resistance and temperature as a function of the applied current density. On (a) we identify three regions where: (i) the magnetization is strongly biased towards the easy-axis in the free layer ($\mathbf{n}_{PMA}$), (ii) there is an auto-oscillation around the easy-axis of the free layer ($\mathbf{n}_{PMA}$), (iii) the magnetization is strongly biased towards the direction of the magnetization in the polarizer ($\mathbf{p}$). The firing behavior correspond to a thermally induced alternation between the auto-oscillation and the strongly biased state as shown on the inset corresponding to the magnetization dynamics of the free layer as shown on the right panel of (a). Notice that the Resistance in this case is given by the $m_x$ component, represented in blue.

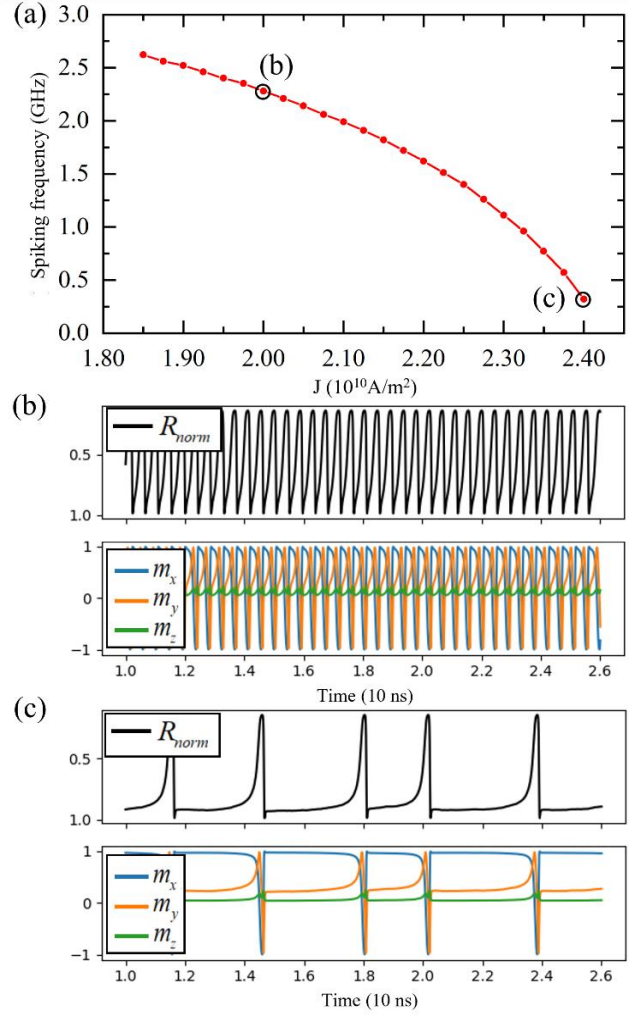

**Figure 3.** (a) Frequency of neuron firing as a function of the applied current density. (b) and (c) show the time domain behavior of the resistance (top), magnetization components (middle, green – $m_z$, orange – $m_y$, blue – $m_x$), and temperature (bottom) for low (2 $10^{10}$ A/m$^2$) and high (2.4 $10^{10}$ A/m$^2$) current density respectively.

Fig. 3(a) shows the spiking frequency as a function of the applied current density emulating the continuous firing under constant input. For low current densities, one observes higher frequencies, Fig 3(b), while the frequency decreases for larger current densities, Fig. 3(c). The current-controlled frequency originates from the non-linear frequency shift linking the frequency and the power of spintronic oscillators as described in previous works and confirmed in several experiments. [81–83] The range of firing frequencies with the experimentally obtained parameters [62] is between 100 MHz and 3 GHz. Moreover, we notice that the period between spikes have small fluctuations associated to temperature induced noise, see that in Fig. 3(c) the time difference between spikes is not constant. The thermal induced stochasticity is important for bio-realistic neural networks. [1,22] The other two thermal phenomenon, i.e., the parameter scaling and the resistance variation driven
9

gain-loss, are responsible for another important feature of the observed dynamics: the characteristic refractory period. The refractory period corresponds to a period succeeding the sharp spike where the output is lower than the average output. The lower output not only deters consecutive spikes as may also be used to tune the synapse weights. The latter allows for the implementation of STDP which is a biologically plausible unsupervised learning mechanism. [57–59] Specifically, the two thermal phenomena generate a temporary stability of the strongly biased state which has a lower resistance than the average resistance and prevents firings for a short period of time.

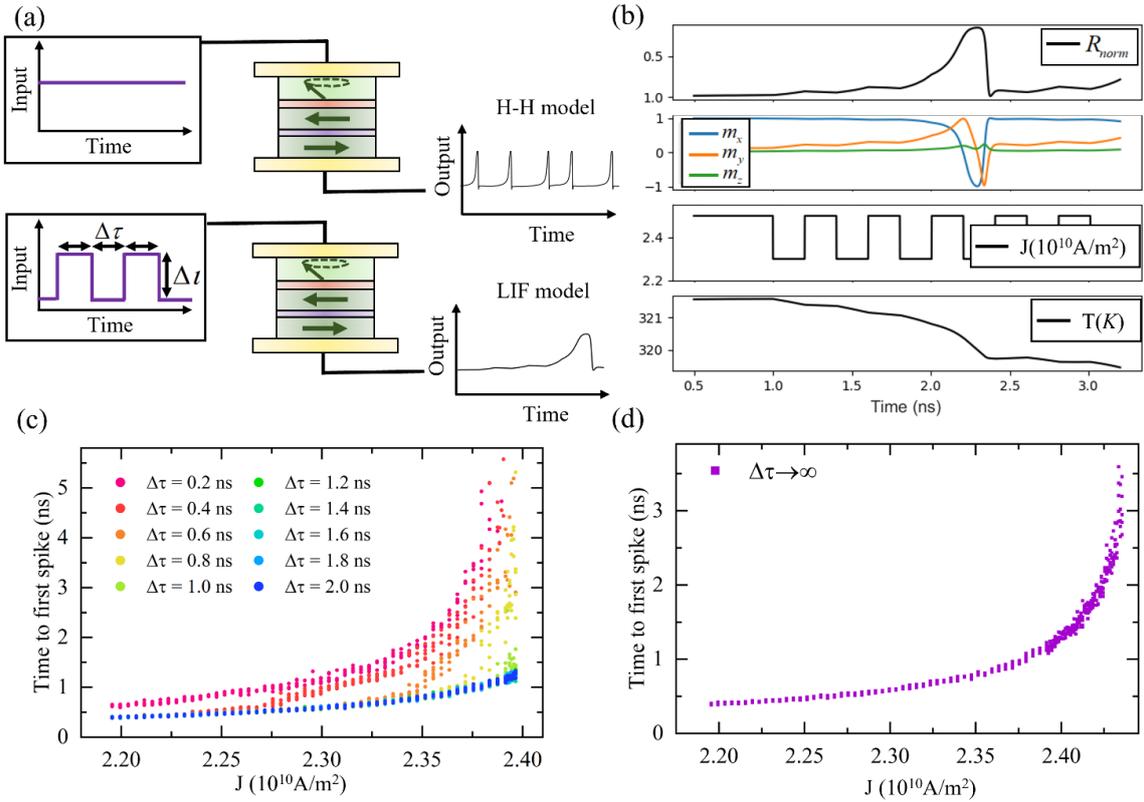

**Figure 4.** (a) Difference between the H-H model and the LIF model. In the LIF model, we consider pulses of amplitude $\Delta\iota$ and period $\Delta\tau$. (b) Evolution of the device properties under the presence of current pulses. From top to bottom panels, we show the normalized resistance, the evolution of the magnetization components, the profile of the applied current and the temperature of the device. (c) shows the time between the first pulse and the first synapse. Different colors represent different pulse sizes, while the x-axis represent the lowest current value in the pulse. To include thermal induced stochasticity, we simulated each pulse amplitude and size five times. (d) The time of the first synapse for a constant current.

Besides emulating the H-H model, the versatile proposed device also works with current pulses within the LIF model. In the latter, pulses generate a time-dependent change in the system which allows to integrate or ignore pulses depending on their amplitude and relative time differences, see Fig. 4(a)- (b). To demonstrate this functionality, we consider the initial/final current as a strong bias current that does not produce spikes and apply a sequence of pulses with a lower current, see Fig. 4(b). In the simulations, we consider both the size $\Delta\tau$ and amplitude $\Delta\iota$ of the pulses to respect to a higher bias current $J_0 = 2.5 \cdot 10^{10}$ A/m$^2$ which shows no spikes. At higher amplitudes $\Delta\iota$, the time to the first spike does not depend on the size of the pulses. To compliment this analysis, we also show in Fig. 4(d) the time of the first spike for constant applied currents, corresponding to a pulse of infinite size. We notice that the time of the first pulse grows almost exponentially until a certain current where firings are no longer expected. Moreover, we notice that for high $\Delta\iota$, i.e. low currents, the time to the first spike becomes rather constant around 0.5 ns due to the existence of the refractory period.

This spiking behaviour, which does not require a resetting mechanism, is rather robust as a function of different parameters and the device properties can be tuned by designing the thickness of the free layer and of the tunnelling layer, which influence the temperature variations.

## IV. ANALOGY BETWEEN HUXLEY-HODGKIN AND LANDAU-LIFSHITZ-GILBERT MODELS

Previous attempts to emulate the H-H model with physical systems rely on coupling several memristors and is not area nor energy efficient. [30,66,84,85] We show, however, the possibility to emulate the H-H model by leveraging the intrinsically nonlinear dynamics of the proposed MTJ device combined with thermal effects. The H-H model is given by,

$$C \frac{dV}{dt} = -F + I, \quad (7)$$

where $V$ is the membrane potential, $C$ is the neuron capacitance, $F$ is the membrane internal current, and $I$ is the sum of external and synaptic currents. The internal current $F$ is a nonlinear function of the potential and three time- and voltage-dependent conductance variables $m$, $l$ and $n$ associated to different biological mechanisms, such as the Potassium, Sodium and leakage currents, [54,66,86]

$$F = g_L(V - V_L) + g_K n^4 (V - V_K) + g_{Na} h m^3 (V - V_{Na}). \quad (8)$$

Here $g_L$, $g_K$, and $g_{Na}$ are constants while $V_L$, $V_K$, and $V_{Na}$ are voltage values associated to each mechanism. The channels $m$, $l$ and $n$ evolve according to first order time differential equations of the form

$$\frac{dx}{dt} = -\frac{1}{\tau_x}(x - x_0(V)), \quad (9)$$

where $x$ stands for $m$, $l$, and $n$. Each channel is characterized by a different characteristic time $\tau_x$, whose different time scales allow for reducing the model with four time-dependent channels to only two. [86,87] We emphasize that the H-H model was obtained phenomenologically to fit experimental data. The model given by Eqs. 7-9 is highly non-linear and cannot be easily simulated by software.

For this reason, there exist several attempts to simplify the model to allow for easier computation. [18,86,87] We show that the magnetization dynamics is equally complex and can emulate the model from Eqs. 7-9 to produce the characteristic sharp spikes followed by the refractory period under constant input based on its own dynamics. We derive a minimal model based on the LLG Eq. 5 to constrain the rich magnetization dynamics to the specific device configuration and to allow for a clear understanding of the model comparability, see Appendix A for details. The minimal model is also useful to reduce the computational cost of numerical simulations. We consider the magnetization in spherical coordinates $\mathbf{m} = \cos\theta(\hat{\mathbf{x}}\cos\varphi + \hat{\mathbf{y}}\sin\varphi) + \hat{\mathbf{z}}\sin\theta$, and by restricting to the dynamics in the self-oscillation regime, with small variations of the polar angle $\theta$, we obtain the following equations of motion,

$$\frac{d\varphi}{dt} = -G + \iota, \tag{10a}$$

where
$$G = \frac{\iota_0}{1 - P(T)^2 \cos\varphi} \sin\varphi - \frac{\gamma\mu_0 M_s(T)}{2}\left(D_x + D_y - 2D_z + (D_x - D_y)\cos 2\varphi\right)\theta - \gamma H_F \cos\varphi\,\theta \tag{10b}$$

with
$$\frac{d\theta}{dt} = -\frac{\iota_0}{1 - P(T)^2 \cos\varphi}\theta\cos\varphi - \frac{\gamma}{2}\left(2H_F\sin\varphi + (D_x - D_y)\mu_0 M_s(T)\sin 2\varphi\right). \tag{10c}$$

Here $\iota_0 = \frac{\gamma\hbar P J_0}{M_s e d}$, and $\iota = \frac{\gamma\hbar P J}{M_s e d}$ are, respectively, the reference spin-current, which bias the magnetization towards **p**, and the spin-current difference that allows for the firing behavior as described in the previous section. We neglected damping because in the active region the magnetic losses are compensated by the negative damping originated by the spin-transfer-torque. Moreover, notice that the model above emulates the H-H behavior even in the absence of the dipolar field, i.e. $H_F = 0$. However, the field is necessary to set the tilted configuration of the magnetization to achieve low critical current for the excitation of self-oscillation of the magnetization. The parameters $M_s(T)$ and $P(T)$ depend on the temperature, as shown in Eq. 3(a)-(b) and evolve according to the time evolution of the of the temperature given by Eq. (2). The model from Eqs. 10(a)-(c), based on the LLG Eq. 5 reproduces the firing behavior reported in Fig. 1(b) and Fig. 2(a) emulating the H-H model. The H-H model is characterized by four dynamical parameters ($V$, $m$, $l$, and $n$), while the LLG equation including thermal effects is described by three dynamical equations for $\theta$, $\varphi$ and $T$, and two equations for the temperature-dependent material parameters $P(T)$ and $M_s(T)$. Thus, the complexity of the MTJ model for the device is comparable to that of the H-H model, allowing the H-H model to be mapped onto the temperature-dependent LLG. In other words, it allows the biological evolution of the membrane potential in the H-H model to be mapped onto the tunable material properties and temperature-dependence of the device. In the proposed single-MTJ hardware implementation, the firing response is fully embedded in the dynamics of the MTJ, and no extra computation is required to simulate the H-H model. In particular, the channels with different time scales in Eq. 9 are mapped to the magnetization dynamics of $\theta$, $\varphi$ and the temperature evolution of the material parameters as discussed in Eqs. 2-4. Hence, the phenomenological constants of the H-H model are replaced by the geometrical and magnetic properties of the MTJ which can be engineered for the hardware requirements of analog SNN implementations. For a comparison between the H-H model and the MTJ device proposed see Table 1 in the introduction.

V. UNSUPERVISED SPIKING NEURAL NETWORK

SNNs are optimal for processing time-dependent data, and the information is encoded in the rate or time difference of spikes. [18,48,88–92] To test the device concept in a simple SNN, we considered a rate-based information encoding. We employed the SNN to perform the classification of binarized images, where pixels can assume values "0" and "1". [93–95] The SNN is given by two layers, an input and output layer, where the MTJs correspond to the firing neurons. For the current input of the first layer, we considered that "0" corresponds to a high current where spikes are not observed, and "1" corresponds to a lower current with a high frequency of spikes. The input current of output neurons $I_{out}$ is given as a weighted voltage of the input neurons $V_{in}$,

$$I_m(t) = \sum_{n=input} \chi_m(t) g(W_{mn} V_n(t)), \quad \text{where} \quad \frac{dg(V)}{dt} = \rho_g \left(V - V_{ref}\right) + \frac{g(V) - g(V_{ref})}{\tau_g}. \quad (11)$$

The function $g(V)$ is an exponential function that allows to integrate spikes over time and maintain the input current for the output neurons rather constant. [96] The parameters $\rho_g$ and $\tau_g$ were chosen according to the range of currents and frequency of spikes desired. The parameter $\chi_m(t) = \{0.1, 1\}$ emulates the lateral inhibition, i.e., if an output neuron spikes the parameter $\chi_m(t)$ is decreased for all other neurons for a period of time given by the characteristic time interval between two spikes of the input neurons. The behavior for a toy-network with two input neurons and two output neurons can be seen in Fig. 5(a). We consider here the full macrospin model given in Eqs. 1-6 and detailed in Appendix A.

For the learning process, we considered an unsupervised mechanism where the weights were updated according to the Hebbian and anti-Hebbian rule. [94,95,97–99] In particular, if an output neuron fired, the weight associated to the input neurons that had just spiked increase while the weight associated to the input neurons that haven't spiked recently decrease. Furthermore, we let the SNN weights evolve until an output neuron has a considerably higher frequency compared to the other neurons. To strengthen the learning, after this process we fix the input currents of the output neurons according to the following algorithm: if the frequency of the output neuron is similar or higher than the frequency obtained with other input images for that same output neuron, the input current is set to the lowest value of the current achieved; else, if the frequency is significantly lower than the frequency obtained with other input images for that same output neuron, the neuron with second highest frequency is set with the lowest current achieved; the input of all other output neurons is fixed not to fire (i.e. the current associated to "0"). With these fixed input currents on inputs and output neurons, the weights then evolve normally according to the Hebbian and anti-Hebbian learning.

We tested the proposed SNN by considering two different sets of three patterns with 3x3 pixels, see Fig. 5(b). The SNN is made of nine input neurons, one for each pixel, and three output neurons, see Fig 5(c). The pixels assumed values "0" or "1", which was uniquely identified to two different currents with different firing rates (0 – low frequency; 1 – high frequency). The tests were performed by initiating the SNN with random weight matrices. After repeating in a temporal sequence each pattern from three to five times, the final weight matrix uniquely identified each of the input patterns, see Fig. 5(d) for an example.

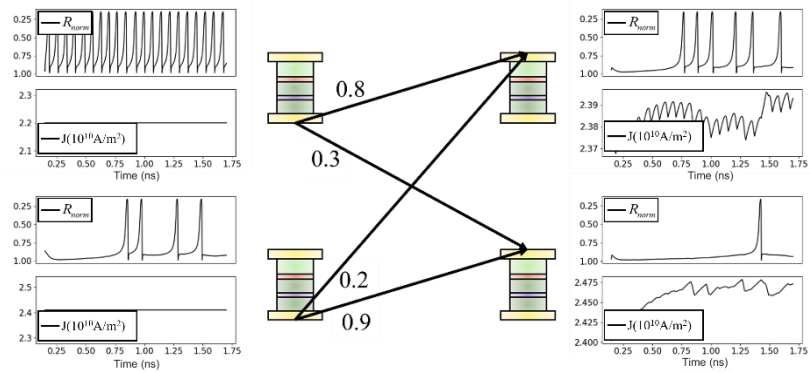

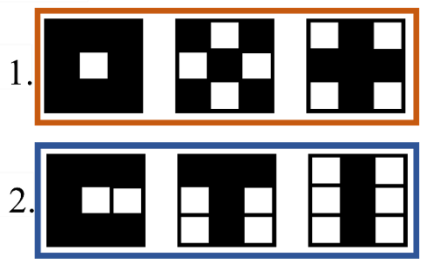

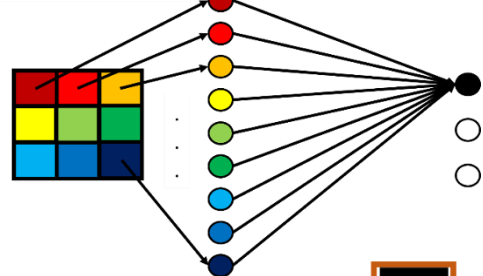

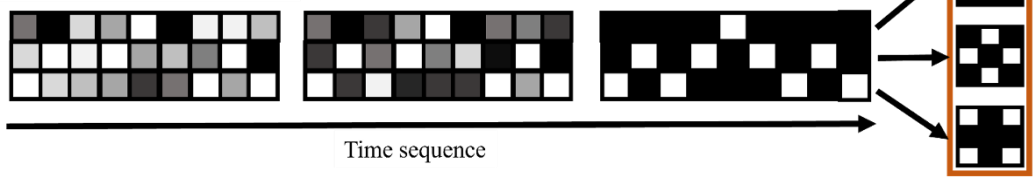

**Figure 5.** An example of SNN built with the proposed MTJ device. (a) shows the behavior of an SNN with 4 neurons. The current of the two input neurons as set as constants. For each MTJ we show the behavior of the resistance (upper panel) and the input current (lower panel). The input currents of the output neurons are generated according to Eq. 11, with the respective weights $W_{mn}$ shown. (b) shows the two sets considered for verifying the learning process and (c) shows a sketch of the Feed-forward All-connected SNN. Each pixel is associated to a single input neuron. (d) shows the evolution of the weight matrix. Starting from a random distribution of weights, after the training process, each figure can be represented by a single output neuron which has the highest frequency.

## VI. SUMMARY AND CONLUSIONS

In this work we have designed a single MTJ device to emulate bio-realistic spiking neurons with physical and geometrical parameters based on an experimental data already published. The device emulates both the simpler LIF model as well as the more realistic H-H model without the need of a resetting mechanism. The proposed device works at room temperature and leverages two physics to produce constant firing at constant input. We demonstrate the frequency dependence on the applied current, as well as a firing mechanism based on the amplitude and length of input pulses. The rate of spikes is in the range from MHz to GHz. The device presents several properties expected for the hardware implementation of bio-realistic neurons, which include: (i) highly scalable, reproducible, and robust; (ii) the characteristic refractory period, showing a depression of the potential after the spike; (iii) a small but non vanishing stochasticity, which allows for random fluctuations without significantly increasing the error rate.

We verified the behavior of the device by simulating a spiking neural network to recognize different figures. The information was encoded in the spiking ratio. The neural network was successfully able to classify the figures. The largest set we considered was 3x3 pixels, due to computational limitations. It is important to emphasize that while these calculations are usually computationally expensive, requiring a significant amount of memory and calculation time, the device's inherent nonlinear and time non-local dynamics can realize the calculation fast at low power input.

Overall, the proposed device corresponds to a low-input, highly reproducible, scalable, robust, CMOS-compatible single MTJ working at room temperature, that emulates the bio-realistic H-H model. The device properties can be engineered to fit the network requirements by modifying the temperature gain/loss of a previously experimentally realized MTJ concept. This proposal allows for an easy drop-in replacement in current SNN CMOS-based hardware implementations to increase area, energy, and memory efficiency.

ACKNOWLEDGEMENTS

The research has been supported by project no. PRIN 2020LWPKH7 funded by the Italian Ministry of University and Research. D.R.R., R. M., M.C., and G.F. are members of the Petaspin team and acknowledge the support from Petaspin association (https://www.petaspin.com/). D.R.R. also acknowledges funding from the Ministerio dell'Università e della Ricerca, Decreto Ministeriale n.1062 del 10/08/2021 (PON Ricerca e Innovazione).

APPENDIX A– MICROMAGNETIC MODEL

In this work, we consider an MTJ experimentally realized in Ref.[[62]] with an MgO barrier sandwiched between a free layer with PMA and an in-plane magnetized fixed layer, see Fig. 1(a) in the main text. The magnetization of the free layer is described by a single magnetization vector which evolves in time according to the Landau-Lifshitz-Gilbert (LLG) Eq. 5 of the main text. The effective magnetic field, $\mathbf{h}_{\text{eff}}$, is given by

$$\mathbf{h}_{\text{eff}} = -\mu_0 M_s \left( D_x m_x \hat{\mathbf{x}} + D_y m_y \hat{\mathbf{y}} + D_z m_z \hat{\mathbf{z}} \right) - H_F \mathbf{n}_F, \quad (A.1)$$

where $\mathbf{m} = \mathbf{M}/M_s$ is unitary magnetization vector of the free layer, $M_s$ is the magnetization saturation, $\mathbf{D}$ is the effective anisotropy vector, [80] which includes the material anisotropy and magnetostatic fields, and $\mu_0$ is the vacuum permeability. We also consider an external field due to the coupling to the fixed layers with strength and direction given by $H_F$, proportional to the magnetization of the fixed layer, and $\mathbf{n}_F$ respectively. The thermal field is modelled by a Gaussian noise with magnitude given by

$$\mathbf{h}_{\text{ther}} = h_{\text{ther}} \mathbf{n}_t, \quad (A.2a)$$

with
$$\left\langle n_{t,i}(t) n_{t,j}(t') \right\rangle = \delta_{ij} \delta(t-t') \quad (A.2b)$$

being the time correlation for the randomized unitary vector $\mathbf{n}_t$, and the magnitude $h_{\text{ther}}$ is given in Eq. 4. The parameters used in the simulations are given on Table A.1.

| Magnetic Parameters | Symbol | Value | Geometric and Thermal Parameters | Symbol | Value |
|---|---|---|---|---|---|
| Saturation Magnetization | $M_s$(T=300 K) | 8.47e5 A/m | Area of Free layer | $A$ | 6.25e4 nm$^2$ |
| Dipolar field from in-plane polarizer and pinning layer | $H_F$ | 7.7 mT | Thickness of Free layer | $d$ | 1.6 nm |
| Polarization constant | $P$(T=300 K) | 0.78 | Curie Temperature | $T_c$ | 800 $K$ |
| Gilbert damping | $\alpha$ | 0.01 | Ambient Temperature | $T_{amb}$ | 300 $K$ |
| Demagnetizing vector | $D$ | (0.1,0.2, -0.95) | Temperature decay time | $\tau_0$ | 4 ns |
| Fixed layer magnetization direction | $\mathbf{n}_F$ | (1,0,0) | Heat efficiency | $\rho/k_B$ | 3.0×10$^{14}$ $K/Js$ |
| | | | Polarization scaling coefficient | $\varepsilon_P$ | 1.5 |

**Table A.1.** Parameters used in the macrospin simulations at room temperature ($T = 300\ K$).

To derive the minimal model from Eqs. 10(a)-(c), we considered the LLG Eq. 5 and substituted the unitary magnetization in the spherical coordinates $\mathbf{m} = \cos\theta(\hat{\mathbf{x}}\cos\varphi + \hat{\mathbf{y}}\sin\varphi) + \hat{\mathbf{z}}\sin\theta$. Notice that for $\mathbf{m}$ parallel to the easy axis, $\theta = \pi/2$, while for $\mathbf{m}$ along $\mathbf{p}$, $\theta = \varphi = 0$. We expanded the LLG equation to linear order in $\theta$, to consider auto-oscillations at high input currents. Thermal effects are included in the minimal model 10(a)-(c) as effective changes to the material parameter values in the LLG Eq. 5, shown in Eqs. 2-4.

Moreover, in the main text, we neglect the contribution of field-like torques. We can include such torques by adding the following term to the LLG Eq. 5,

$$\tau_{FL} = \alpha_{fl}\tau_{STT}\mathbf{m}\times\mathbf{p}, \qquad (A.3)$$

where $\alpha_{fl}$ is a coefficient for the strength of the field-like torque and is often assumed to be small (i.e. $\alpha_{fl} \ll 1$) [100]. Fig. A.1 shows the spiking behavior for different currents and $\alpha_{fl}$. We notice that the field-like torque qualitatively increases the frequency of the spikes, requiring a higher current for a lower spiking frequency.

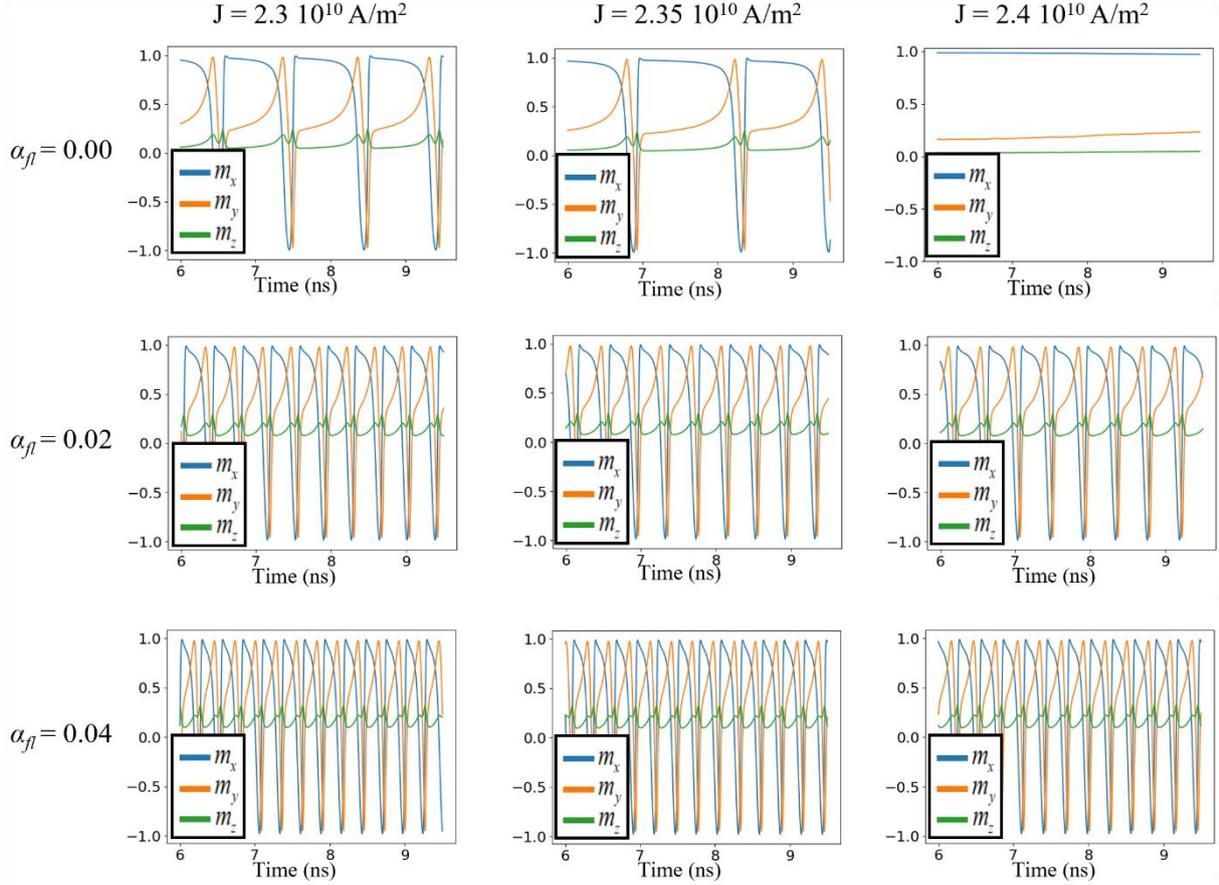

**Figure A.1:** The behavior of the magnetization dynamics for different currents and including the field-like-torque. The material parameters are the ones from Table A.1 in the manuscript.